\documentclass[a4paper]{article}

\usepackage{arxiv}

\usepackage[utf8]{inputenc} %
\usepackage[T1]{fontenc}    %
\usepackage{hyperref}       %
\usepackage{url}            %
\usepackage{booktabs}       %
\usepackage{amsfonts}       %
\usepackage{nicefrac}       %
\usepackage{microtype}      %
\usepackage{lipsum}
\usepackage{cite}

\usepackage{textgreek}
\usepackage{graphicx}
\graphicspath{ {./images/} }
\usepackage{amsmath}

\usepackage{xcolor}
\usepackage{soul}

\usepackage[capitalise, noabbrev]{cleveref}

\title{A RANK-3 NETWORK REPRESENTATION FOR SINGLE-AFFILIATION SYSTEMS}

\author{
  Alexander O.~Hultin\\
  Department of Mechanical Engineering\\
  University of Bath\\
  Bath, UK \\
  \texttt{A.E.O.Hultin@bath.ac.uk} \\
   \And
  James A. Gopsill\\
  Department of Mechanical Engineering\\
  University of Bath\\
  Bath, UK \\
  \texttt{J.A.Gopsill@bath.ac.uk} \\
   \And
  Nigel Johnston\\
  Department of Mechanical Engineering\\
  University of Bath\\
  Bath, UK \\
  \texttt{D.N.Johnston@bath.ac.uk} \\
   \And
  Linda B. Newnes\\
  Department of Mechanical Engineering\\
  University of Bath\\
  Bath, UK \\
  \texttt{L.B.Newnes@bath.ac.uk} \\
}

\begin{document}
\maketitle \raggedbottom

\begin{abstract}

Single-affiliation systems are observed across nature and society.
Examples include collaboration, organisational affiliations, and trade-blocs. 
The study of such systems is commonly approached through network analysis.
Multilayer networks extend the representation of network analysis to include more information through increased dimensionality. Thus, they are able to more accurately represent the systems they are modelling. However, multilayer networks are often represented by rank-4 adjacency tensors, resulting in a $N^2M^2$ solution space. Single-affiliation systems are unable to occupy the full extent of this space leading to sparse data where it is difficult to attain statistical confidence through subsequent analysis. To overcome these limitations, this paper presents a rank-3 tensor representation for single-affiliation systems. 
The representations is able to maintain full information of single-affiliation networks in directionless networks, maintain near full information in directed networks, reduce the solution space it resides in ($N^2M$) leading to statistically significant findings, and maintain the analytical capability of multilayer approaches. 
This is shown through a comparison of the rank-3 and rank-4 representations which is performed on two datasets: the University of Bath departmental journal co-authorship 2000-2017 and an Erdős–Rényi network with random single-affiliation. 
The results demonstrate that the structure of the network is maintained through both representations, while the rank-3 representation provides greater statistical confidence in node-based measures, and can readily show inter- and intra-affiliation dynamics.

\end{abstract}


\section{Introduction}

Affiliation systems are observed across nature and society. These systems feature a set of system elements, such as people, animals, organisations, and countries, that are affiliated to one or more affiliations, such as teams, social groups, alliances, and trade blocs. Examples include, Zachary's Karate Club, board of directors, socioeconomic class, organisational coordination, and trade~\cite{Newman2004b,Faust1997,Cote2017,Chen2013,Kohl2014}. An affiliation is defined herein as links between an object to another set of objects. This can be how a word is associated with a word class (e.g.\ verb, noun, adjective), how a person is connected with an organisation, or how an airline is part of an alliance. Affiliation systems are defined herein as systems of interconnected elements where each element is associated with one or more affiliations. Understanding the structure and dynamics of such systems and how affiliations affects system behaviour enables stakeholders to make informed decisions. 

Single-affiliation systems are a subset where system elements are singularly affiliated.  Such systems exist naturally in society, as exemplified in Zachary's Karate Club and political party membership~\cite{Newman2004b,Porter2005}. These systems often have a significant impact on society. For instance, interdisciplinary research has been identified as being vital to addressing real world problems, whilst trade-blocs such as customs unions affect supply-chains and negotiations~\cite{Davidson2015,Kohl2014}. As single affiliations are exclusive (i.e. either a node belongs or it does not), they serve as boundaries and the inter/intra-affiliation dynamics related to these affiliations is a topic of interest. 

The study of similar systems has been commonly approached through network analysis~\cite{NewmanBook,BarabasiBook}. Networks are defined here as a set of nodes connected by a set of links. All nodes and links are of a given class respectively.~\cite{NewmanBook}. That is to say that nodes and links are respectively homogeneous. 

Key to the successful application of network analysis is the ability to represent the system effectively so that meaning can be assigned to the findings from the analyses performed and underlying dynamics of the systems can be explored. It has been used to investigate many phenomena such as transport, knowledge creation, document co-authorship, social networks, and trade~\cite{Newman2004a,Newman2001,Newman2004,Furusawa2007}.  

However, limitation of homogeneity in nodes and links is a limiting factor and can oversimplify system representations. For instance, the interdependence of electrical and computer networks caused a cascading failure resulting in large scale blackout in Italy in 2003 \cite{Buldyrev2010}. In traditional networks, it would be impossible to model this. To overcome this, the field of multilayer networks has gained traction. Multilayer networks exist in many different forms, but are starting to form analytical norms. One of those norms is to represent multilayer networks through a rank-4 adjacency tensor \cite{DeDomenico2013}. This form occupies a solution space of $N \times N \times M \times M$ (where $N$ is the number of nodes and $M$ is the number of layers). This dimensionality can quickly cause datapoints to be sparsely represented. This is of particular importance to single-affiliation systems, where only $N(N-1)$ links can exist.

To overcome the limitations of current approaches, this paper proposes a rank-3 tensor multilayer network representation for single-affiliation systems that maintains the full structural information of a rank-4 multilayer undirected network, and near full structural information in directed networks. It does so by reducing the dimensionality of the representation. Achieving this enables researchers to continue to study large single-affiliation systems with statistical confidence. 

The paper continues by discussing the related work in affiliation systems and the network approaches that have been developed to study them (Section 2). Through this, the challenge of dimensionality and gap in the study of single-affiliation systems are identified. Section 3 introduces the rank-3 tensor model for single-affiliation systems. A method to evaluate and compare to a rank-4 tensor model applied to a University departmental co-authorship dataset and Erdős–Rényi network with random single-affiliation is described (Section 4), applied, and discussed (Section 5). The implications are then concluded with the key contributions (Section 6).

\section{Relevant works}

This section reviews multilayer network analysis to understand the benefits and drawbacks of these approaches to representing single-affiliation systems.

Multilayer Networks have their roots in sociology where it was identified that different types of social relationships are important to consider and that approximating the different relationships as being equal was an inadequate approach~\cite{Krackhardt1987,Padgett1993,Wasserman1994}. 
Whilst traditional network analyses have yielded vital insights regarding networks such as the small-world and scalefree properties, the assumption that the nodes and links have to be of a single type limits further investigation~\cite{Watts1998,Barabasi1999}. As discussed, this limitation is exemplified in cascading failures of interdependent electrical and internet networks that led to a blackout in Italy in 2003 \cite{Buldyrev2010}.

\begin{figure}
    \centering
    \includegraphics[width=0.66\textwidth]{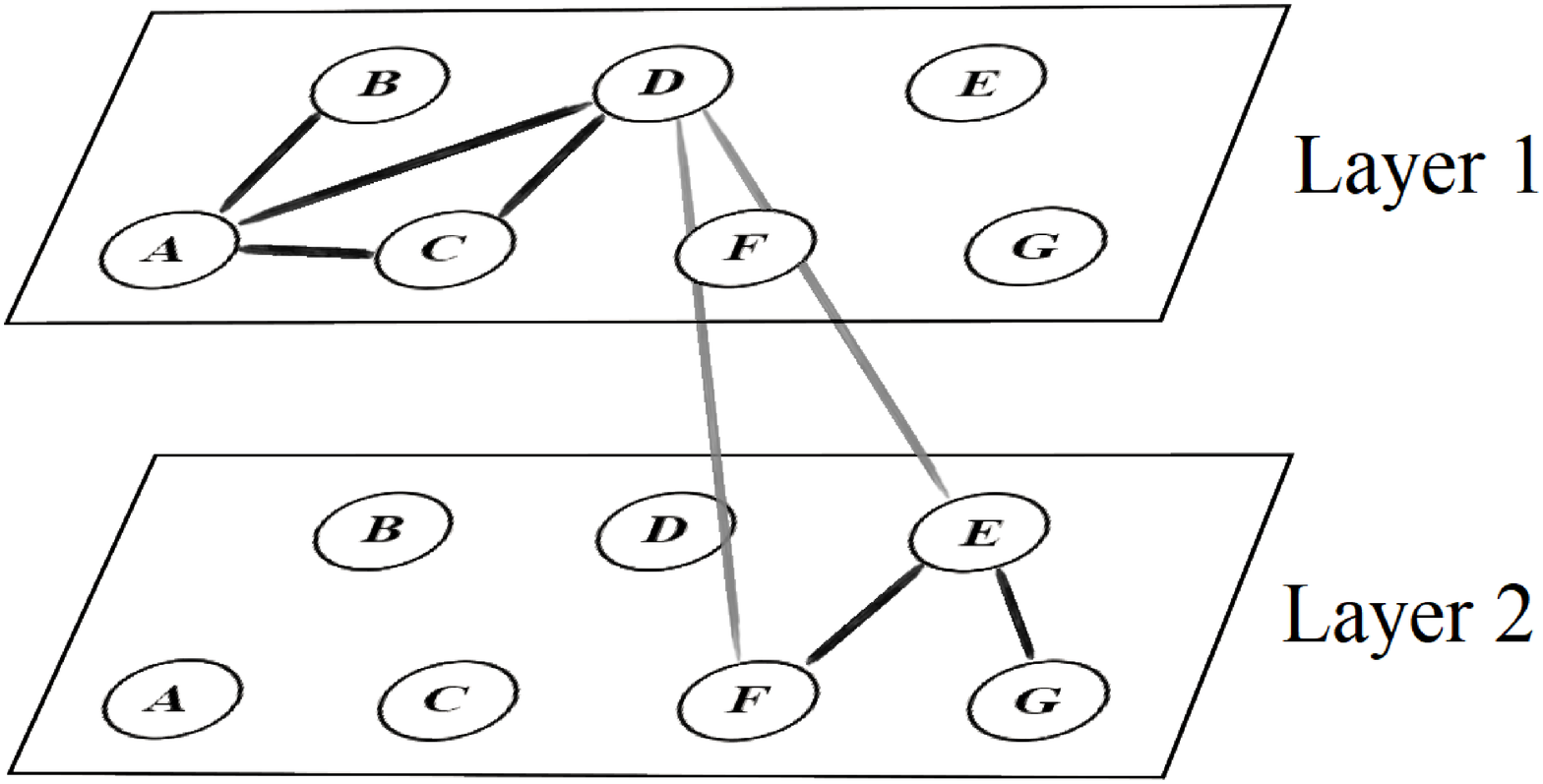}
    \caption{An exemplar multilayer network. This consists of 7 nodes labelled A, B, C, D, E, and F, each of which exists in layers 1 and 2. In each layer, the set of links are different and may represent different types of links (e.g. in transport networks, where each node is a location, the links could be train, airplane, and bus links between the locations. Links may also exist between layers (e.g. collaborations occurring between different departments), exemplified by the links (D,E) and (D,F).}
    \label{fig:fig1}
\end{figure}

As such, multilayer networks are the study of networks wherein nodes can connect to nodes on other layers. 
These layers can provide the additional information required to represent different types of node and link.
Thus, a definition of multilayer networks is a network with multiple layers with a known set or sets of nodes connected by sets of links within and between layers. In this research, all layers are node-aligned (i.e.\ all nodes exist in all layers) where the nodes form connections with any other node in any other layer. 
This is shown conceptually in Figure~\ref{fig:fig1}. 
Such representations have been used to improve our understanding of urban infrastructure, transport, text, collaboration, social media networks, terrorist networks, email networks, trade, and zoology \cite{Zhao2016,Rombach2017,Cozzo2015,Cardillo2013,Cardillo2013a,Nicosia2015,Li2012, Sun2009,Ng2011,Coscia2013,Berlingerio2013,Berlingerio2013a,Menichetti2014,Battiston2016,Ghariblou2017,wang2017strong,Halu2013,Iacovacci2016,Starnini2019, Barrett2012}. 

Formally, a multilayer network is defined by $\mathcal{G} (V, E, L)$ where:

\begin{itemize}
    \item $V=\{n_1,n_2,\ldots,n_N\}$ is the set of $N$ nodes representing the system elements;
    \item $L=\{l_1,l_2,\ldots,l_N\}$ is the set of $M$ layers representing affiliations; and,
    \item $E=\{e_{ij\alpha\beta}\}$ is the set of links relating nodes across and within layers.
\end{itemize}

It has been shown that multilayer networks can be represented as an adjacency tensor, allowing tensor operations to be performed on the network broadening the analytical approaches that can be applied \cite{DeDomenico2013}. 
They are able to represent phenomena that other methods are unable to, such as emergent diffusion behaviour~\cite{Gomez2013, SoleRibalta2013}. 
It is also possible to use Higher-Order Singular Value Decomposition (HOSVD) to create a multilayer centrality measure~\cite{Kleinberg1999}, or to use similar techniques to create a higher-order modularity analysis~\cite{Dunlavy2011,Bonacina2015}. 
This creates opportunities to identify important systems elements across a wider system environment. 
For instance, in a case such as the interdependence of the electrical grid and internet communication network, it could help use identify system critical elements that require systemic fail-safes, or identify cluster at risk of cascading failures that are not evident in any one network.

An adjacency tensor suitable for affiliation systems is a rank-4 tensor of the format: $\mathcal{A} \in \{0,1\}^{N \times N \times M \times M}$. 
This research uses the notation where the tensor element, $\mathcal{A}_{ij \alpha\beta}$ refers to the adjacency of node $i$ in layer $\alpha$ to node $j$ in layer $\beta$, such that

\begin{equation}
    \mathcal{A}_{ij\alpha \beta} =\Bigg \{ 
    \begin{matrix}
        1 & \text{if} & ((i, \alpha), (j,\beta) \in E \\
        0 & \text{if} & ((i, \alpha), (j,\beta) \notin E
    \end{matrix}
\end{equation}

For clarity, tensors are represented in math calligraphy (e.g. $\mathcal{X}$). Subscripts using Latin letters denote nodes, whilst Greek letters are used to denote layers. 
Where there are two letters of the same type, a relationship is being described in the direction of source to target.

However, note that when representing links in an adjacency tensor, an equivalency is being drawn. 
That is to say, the elements of the tensor show that there exists a link between nodes and layers, and treats these as being equal (or in the case of weighted adjacency, as being on the same scale). 
This research adopts the stance that even in tensors, the links must be of the same class and that different types of links cannot be represented in the same adjacency tensor. 
If inter-layer links exist, these must be the same type of links within layers.

Multilayer networks can be applied to affiliation systems and model system elements inter-connectivity. This approach represents each system element as a node, each relationship between system elements as a link, and each affiliation as a layer. This adheres to the definition of a multilayer network, $\mathcal{G}$ and can represent the system intra-affiliation dynamics (when $\mathcal{A}_{\alpha\beta}: \; \alpha=\beta$) and in inter-affiliation dynamics (when $\mathcal{A}_{\alpha\beta}: \; \alpha \neq \beta$).

However, in single-affiliation systems, wherein a node can only have a single affiliation, an important limitations arises. Consider that the system that we are representing consists of inter-connected system elements where all elements have an affiliation. 
The maximum number of possible links a system element can have is to all other system elements, i.e. $N-1$. 
As there are $N$ elements, then there are only $N(N-1)$ possible links. 
As the adjacency tensor consists of $N^{2}M^{2}$ elements, the maximum possible data density can only be $\sim 1/M^{2}$ in fully connected systems and will in reality be significantly more sparse.

This limitation becomes even more apparent with increasing $M$ limiting rank-4 approaches in studying large single-affiliation networks due to its sparse representation that limits the ability gain statistical confidence in any analyses performed.

\subsection{Summary}

Multilayer networks approaches have been used successfully in the study of affiliation systems. Multilayer networks exist in many different forms, but represent the interconnectivity between the same set of nodes well. However, limitations arise when studying single-affiliation systems. The majority of applied cases have represented different types of relationships as opposed to affiliations, whilst the dimensionality can be an issue, particularly in empirical datasets that can have many layers and few datapoints. To overcome the limitation in current representations, this paper presents a rank-3 tensor network representation for single-affiliation systems.

\section{A rank-3 representation of single-affiliation systems}

\begin{figure}
  \centering
  \includegraphics[width=0.75\textwidth]{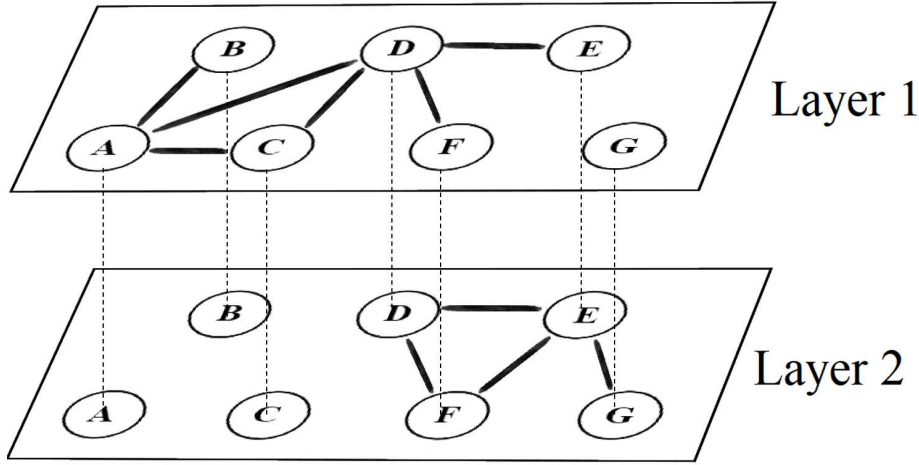}
  \caption{An exemplar rank-3 multiplex network. This consists of 7 nodes labelled A, B, C, D, E, F, and G. A, B, C, and D are affiliated with Layer 1, and E, F, G are affiliated with Layer 2. Layers are defined by affiliation (e.g. Layer 1). Each node has their full personal network represented in their affiliation layer. This provides higher-level view of the different interactions. It can be seen that node D has three links in Layer 1 and two links in Layer 2. This shows that two links, (D,E) and (D,F), are inter-affiliation (as exhibited by the link overlap). All other links are intra-affiliation. The dashed lines exemplify the self-connectivity between layers, which is necessary to establish connectivity between layers.}
  \label{fig:fig3}
\end{figure}

This paper proposes that a rank-3 tensor representation is a more appropriate and capable model of reflecting single-affiliation systems. 
Such a representation can be achieved if the inter-affiliation links in a rank-4 multilayer network are represented within the source and target affiliation layers (i.e.\ when $\alpha = \beta$) as oppsed to between them. This results in inter-layer links being represented twice.  
As there can only be a single connection between two nodes, instances where there is link overlap (i.e.\ same link existing in two different layers) must now be an inter-affiliation. 
This is shown conceptually in Figure \ref{fig:fig3}. 
The transform is given by

\begin{equation}
    {\mathcal{A}_{3}}_{\alpha} = \sum_{\beta}^{M} {\mathcal{A}_{4}}_{\alpha \beta}
\end{equation}

Where the tensor for the rank-3 representation is denoted by ${\mathcal{A}_{3}}$ and the rank-4 representation is denoted by ${\mathcal{A}_{4}}$. Note that the resulting tensor, ${\mathcal{A}_{3}}$, all elements have a value of $0$ or $1$ as there are no link overlaps in the multilayer representation, ${\mathcal{A}_{4}}$. 

Upon observation, it may seem that the rank-3 tensor loses the ability to effectively investigate inter-affiliation links. 
However, inter-affiliation links are represented in two manners in the rank-3 tensors. 
Firstly, the affiliation of a node can be deduced by virtue of its full personal network being exhibited in a single affiliation. It is almost always possible to determine affiliation, with the exception of a special case. The special case where a node's only links are to nodes from another affiliation, and that the neighbours' affiliation are equally indeterminate. This would require the full chain of networks being indeterminate.
Secondly, inter-affiliation links exist on two layers: in the source node affiliation layer and in the target node affiliation layer. 
Intra-affiliation links can only exist in the affiliation layer. 
Thus, it is possible to discern structurally whether a link is intra-affiliation or inter-affiliation if there is link overlap in two layers.

Using this, it is possible to fully recreate the undirected rank-4 multilayer network through the following transform
 
 \begin{equation}
    {\mathcal{A}_{4}}_{ij \alpha \beta}=
    \begin{cases}
    \begin{matrix}
        1 & \text{if} & \mathcal{A}_{3_{ij \alpha}} / \sum_{\gamma}^{M} \mathcal{A}_{3_{ij \gamma}} = 1 & | & \alpha = \beta \\
        1 & \text{if} & \mathcal{A}_{3_{ij \alpha}} + \mathcal{A}_{3_{ij \beta}} = 2 & | & \alpha \neq \beta \\
        0 & \text{if} & \mathcal{A}_{3_{ij \alpha}} + \mathcal{A}_{3_{ij \beta}} = 1 & | & \alpha \neq \beta \\
        0 & \text{if} & \mathcal{A}_{3_{ij \alpha}} + \mathcal{A}_{3_{ij \beta}} = 0 &   & 
    \end{matrix}
    \end{cases}
\end{equation}
 
 For directed networks, it is possible to deduce the adjacency tensor nearly in the same way. 
 The difference lies in that it is not possible to directly find the order of the affiliations (i.e. $\alpha\beta$ or $\beta\alpha$). 
 However, if the nodes' affiliations can be inferred as previously described, then the order is trivial to infer. 
 
 As such, this rank-3 tensor representation maintains the full set of structural information from the rank-4 framework whilst reducing the problem space from $N^2M^2$ to $N^2M$. 
 Furthermore, the rank-3 representation is more efficient with data storage as rank-4 representation is explicit and cannot have link overlap. 
 The amount of information contained in both representations is the same: a node can be linked to another node, which means there are $N(N-1)$ possible links spread over $N^2M^2$ or $N^2M$ data space for the rank-4 and rank-3 representations respectively. 
 Thus, the rank-4 representation can only utilise $1/M^2$ of the data space whilst the rank-3 representation can use $1/M$ of the data space.

\section{Comparing rank-3 and rank-4 multilayer networks for single-affiliation systems}

Two multilayer representations for affiliation networks have been examined. For clarity, the multilayer approach is referred to as the rank-4 representation, whilst the multiplex approach is referred to as the rank-3 representation. 

The rank-4 representation provides a highly granular view of the network, whilst the rank-3 representation contains the same structural information in a reduced dimensionality at the cost of specificity. That is to say that if the purpose of the affiliation network is to investigate the interaction between two specific layers, there is no benefit to the rank-3 representation. If on the other hand the purpose of the research is to understand which affiliation is the most central (e.g. in trade, which trade bloc is the most central, or which country trades the most outside its bloc), the rank-3 representation is appropriate and provides benefits. In the study of intra- and inter-affiliation structures and dynamics, both representations are capable of identifying pertinent links. 

However, in the endeavour of choosing a representation it is necessary to understand where the differences, similarities, benefits, and drawbacks arise. As such, this section compares the representations. It does so by first adopting a series of metrics to represent the structures, and then comparing these metrics in both representations on the University of Bath journal co-authorship 2000-2017 dataset and a randomly generated multiplex network.

\subsection{Datasets}

To demonstrate the capability of the rank-3 representation for single-affiliation systems, a comparison with a rank-4 representation on a University departmental co-authorship dataset and Erdős–Rényi network with random single-affiliation.

The University of Bath journal co-authorship is a single-affiliation system where each author is affiliated with a department. 
The dataset consists of all journal publications from 2000-2017 and was pulled from the University PURE repository. 
The dataset features 23,468 papers, 2,187 unique authors from the University of Bath, 6,578 co-authorship relationships, and 17 departments. 
The dataset also included operational departments, such as the Vice-Chancellor's Office, as well as unaffiliated authors, which were omitted from the analysis.

\begin{table}[h!]
\centering
\caption{Table outlining the number of authors affiliated with specific departments.}
\begin{tabular}{||l c||} 
 \hline
 Departments & Number of affiliated authors \\ [0.5ex] 
 \hline\hline
Biology and Biochemistry & 231 \\
\hline
Chemistry & 373 \\
\hline
Social and Policy Sciences & 53 \\
\hline
Physics & 145 \\
\hline
Chemical Engineering & 114 \\
\hline
Politics Languages and International Studies & 19 \\
\hline
Health & 124 \\
\hline
Economics & 21 \\
\hline
Psychology & 90 \\
\hline
Education & 34 \\
\hline
Mechanical Engineering & 267 \\
\hline
Mathematical Sciences & 99 \\
\hline
Pharmacy and Pharmacology & 200 \\
\hline
Architecture and Civil Engineering & 117 \\
\hline
Electronic and Electrical Engineering & 124 \\
\hline
School of Management & 74\\
\hline
Computer Science & 68 \\
 \hline
\end{tabular}
\label{table:dataset}
\end{table}

An Erdős–Rényi network with random single-affiliation was also generated. The network consisted of 2,000 nodes with a link attachment probability of 0.003 in order to randomly distribute $\sim$6,000 links to reflect the co-authorship network. The nodes were randomly assigned one of ten affiliations. For the rank-3 representation this populates $\sim 1.5\times10^{-4}$ data points per tensor element, whereas the rank-4 representation populates $\sim 1.5\times10^{-5}$ data points per tensor element.

\subsection{Comparison metrics}

Rather than trying to create a new measure akin to a multilayer degree (as some studies have done \cite{Lytras2010,Brodka2012,Berlingerio2011,Berlingerio2013,DeDomenico2014,Menichetti2014}), this paper takes the approach that several different measures capture the whole more effectively. Three metrics are adopted that adequately capture the intricacies of affiliation network structures. These measure are not intended to be exhaustive, but rather provide a means to compare the rank-3 framework to the rank-4 representation so that the benefits, drawbacks, similarities, and differences of both can be highlighted and discussed. 

In order to compare the rank-3 and rank-4 representations, it is necessary to choose metrics that can be applied to both. Given that single-affiliation systems are focused on connectivity between nodes of different affiliations, it is natural to create measures that focus on this connectivity. Such connectivity occurs across $N \times N$ dimensionality. For the rank-3 representation, there are $M$ such occurrences, whereas for the rank-4 representation, there are $M^2$ such occurrences. These occurrences are referred to as \textbf{slices}. Rather than write this twice for each representation, a slice, $\lambda$, is defined. 
For the rank-3 representation: $\lambda \in \{\alpha \in M\}$. For the rank-4 representation: $\lambda \in \{(\alpha\beta) \in M^{2}\}$, where $\mathcal{A}_{\lambda} \in \{0,1\}^{|N| \times |N|}$.

Using these slices, three metrics have been applied.

\begin{enumerate}
    \item Degree centrality
    \item Node activity 
    \item Slice-Pair Closeness
\end{enumerate}

\subsubsection{Degree distribution}

The first measure is the degree distribution that provides us with an insight into the structure that is captured by the rank-3 and rank-4 representations~\cite{Albert2002} and is given by

\begin{equation}
    k_{i \lambda} = \frac{1}{N} \sum_{j}^N \mathcal{A}_{ij \lambda}
\end{equation}
    
This measures the degree of node $i$ in slice $\lambda$. As such, $k_{\lambda}$ is a vector of length $N$. There is one such vector for each slice. Each slice can then produce a degree distribution, $P(k_{\lambda})$. For empirical data, a scalefree distribution is expected \cite{Barabasi1999,Albert2002,BarabasiBook}, the degree distribution can be approximated by

\begin{equation}
    P(k_{\lambda}) \sim k^{-\gamma_{\lambda}}
\end{equation}

Where $\gamma_{\lambda}$ is an exponent that approximates the distribution of degrees in slice $\lambda$. This can be used to determine statistical significance, where a distribution is approximated by an equation of this format and a linear regression can performed to establish whether it is significant to the 0.05 threshold. 

For random graphs, a Poisson distribution is expected and the degree distribution can be approximated by

\begin{equation}
    P(k_{\lambda})=
        \begin{pmatrix}
            N-1 \\
            k_{\lambda} 
        \end{pmatrix}
    p^{k_{\lambda}}(1-p)^{N-1-k_{\lambda}}
\end{equation}

Where $p$ is the probability of a link existing. The exponent for scalefree networks and the mean degree for random graphs can serve as a characteristic measure to compare slices. 

Therefore, the degree distribution produces two important results: it enables the statisical significance to be established within each representation and allows us to compare the behaviour within and between layers.

\subsubsection{Node activity distribution}

The second measure is node activity distribution, which analyses the structure across the slices~\cite{Nicosia2015,Battiston2014}. 
This measures how many affiliations a node is active in and the distribution can give an indication of inter-affiliation structure of the system.
In this respect, it is analogous to the degree centrality across the layers.
    
A node is said to be active in a slice if it has a link in that slice. This is defined as

\begin{equation}
    b_{i \lambda}
    \left \{
    \begin{matrix}
        1 & if & k_{i \lambda}>0 \\
        0 & if & k_{i \lambda}=0 
    \end{matrix}
    \right .
\end{equation}

The node activity is then given by

\begin{equation}
    B_{i} = \frac{1}{\bar{M}} \sum_{l}^{\bar{M}} b_{i \lambda}
\end{equation}

Where $\bar{M}$ is $M$ for the rank-3 framework and $M^{2}$ for the rank-4 representation. 
This creates a scalar value for each node $i$, resulting in a vector of size $N$. 
This can then be used to create a probability distribution, $P(B)$, which can provide information regarding the inter-affiliation structure of the network.
    
\subsubsection{Slice-Pair Closeness distribution}

The third measure is slice-pair closeness, which analyses the structure of slices and how they relate to each other~\cite{Nicosia2015}. 
By comparing if a node is active in a pair of slices, it creates a node-centric way to determine the similarity between slices. 
The slice-pair closeness is defined as
    
\begin{equation}
    Q_{\lambda_{x}\lambda_{y}} = \frac{1}{N} \sum_{i}^{N} b_{i \lambda_{x}} \cdot b_{i \lambda_{y}}
\end{equation}
    
Where $\lambda_{x}$ and $\lambda_{y}$ are two known slices. 
This creates $M^{2}$ values for the rank-3 framework and $M^{4}$ values in for the rank-4 framework. 
This could be aggregated to a slice closeness centrality given by
    
\begin{equation}
    Q_{\lambda_{x}} = \frac{1}{\bar{M}} \sum_{\lambda_{y}}^{M} Q_{\lambda_{x} \lambda_{y}}
\end{equation}
    
This would provide a way to determine which slice is the most similar to all the other slices. 
These can be used to create probability distributions, $P(Q_{\lambda_{x}\lambda_{y})}$ and $P(Q_{\lambda_{x}})$, which can provide useful information regarding the structure of how affiliations interact with one another. 

\section{Results}

The previous section established how it is that the two representations can be compared, the metrics to compare them, and the two datasets analysed for the comparison: the University of Bath author department-affiliation journal co-authorship 2000-2017, and a randomly generated graph with randomly assigned affiliations. This section applies these metrics to the two datasets. 

\subsection{Degree distribution}

The degree distribution provides insight into the network structure in specific slices. 
In the rank-4 representation it provides the structure in every type of affiliation interaction possible. However, as the dimensionality is significant, many of the slices cannot produce any statistically significant trend. 
The rank-3 representation represents the structure for every individual affiliation. 
However, this does not provide information on specific interactions between affiliations. 

\textbf{For the empirical dataset}, a larger $\gamma$ in $P(k)\sim k^{-\gamma}$ usually occurs when there are many nodes that are poorly connected, whereas lower $\gamma$ values usually occur when there is less inequality of connectivity and there is a greater proportion of nodes with more links.

\begin{figure}
  \centering
  \includegraphics[width=1\textwidth]{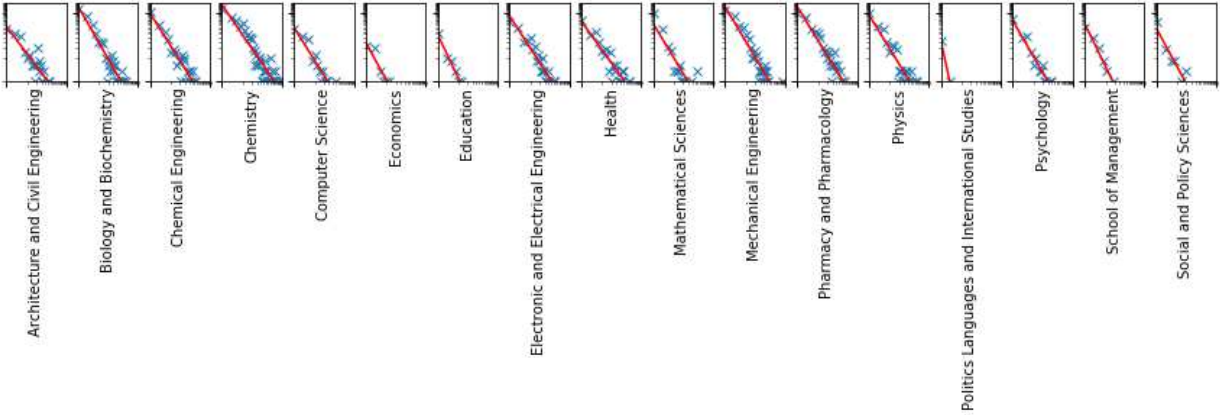}
  \caption{Thumbnails the degree distributions for each slice in the University of Bath co-authorship dataset using the rank-3 framework. The crosses represent the data points and the line represents the trend line approximated by $P(k_{\lambda}) \sim k^{-\gamma_{\lambda}}$. There is no trend line in distributions that do not meet the 0.05 statistical significance threshold.}
  \label{fig:fig4}
\end{figure}

\begin{figure}
  \centering
  \includegraphics[width=1\textwidth]{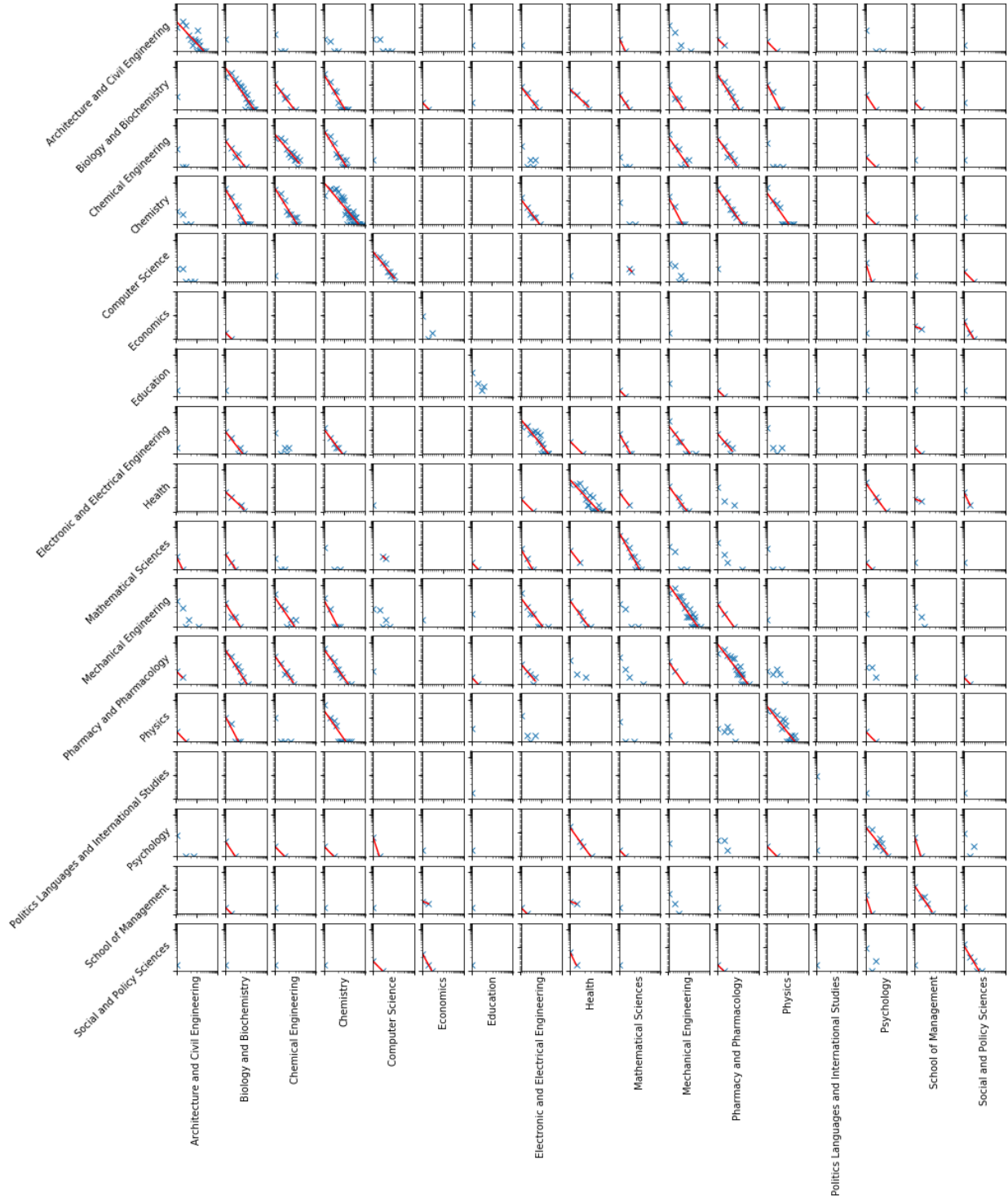}
  \caption{Thumbnails the degree distributions for each slice in the University of Bath co-authorship dataset using the rank-4 framework. The crosses represent the data points and the line represents the trend line approximated by $P(k_{\lambda}) \sim k^{-\gamma_{\lambda}}$. There is no trend line in distributions that do not meet the 0.05 statistical significance threshold.}
  \label{fig:fig5}
\end{figure}

Thumbnails of the resulting degree distributions for departmental affiliation are shown in Figure \ref{fig:fig4} and Figure \ref{fig:fig5} for the rank-3 and rank-4 frameworks respectively. 
In cases where statistically significant trends can be drawn, the scalefree property emerges in both the rank-3 and rank-4 slices. 
However, as these thumbnails demonstrate, the possible data space is much higher in the rank-4 framework, making the data more sparse. 
27.7\% of slices in the rank-4 are statistically significant whilst 94.1\% are statistically significant in the rank-3.

\begin{figure}
  \centering
  \includegraphics[width=0.50\textwidth]{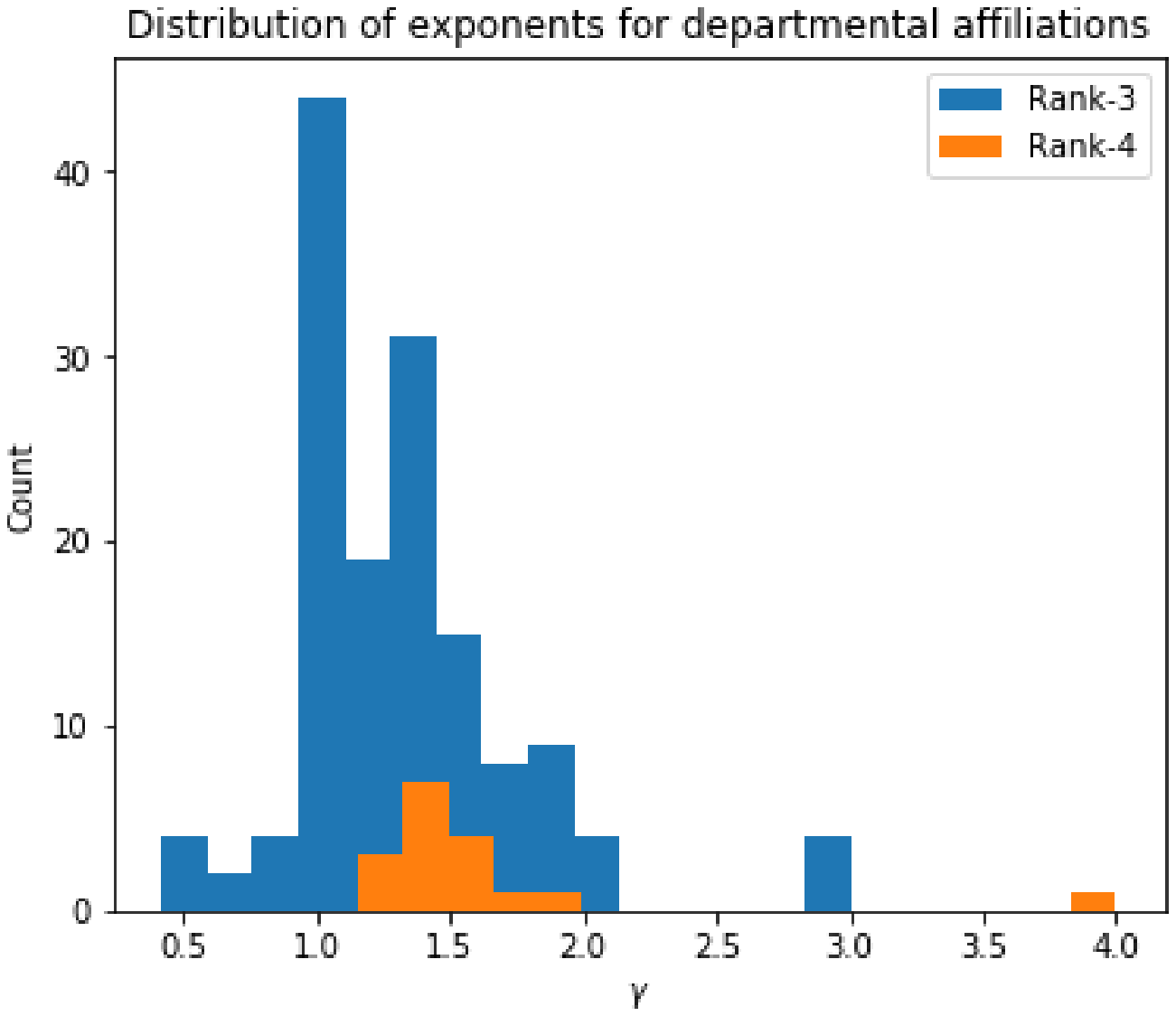}
  \caption{Histograms of the approximated exponents of the slice degree distributions for the rank-3 and rank-4 representations of the University of Bath co-authorship dataset. These only include trend lines for statistically significant regressions. The bins are selected based on the Freedman-Diaconis rule. The shapes and peak values do not differ significantly between the rank-3 and rank-4 frameworks.}
  \label{fig:fig6}
\end{figure}

Approximating the degree distributions by their exponents, the distribution of the exponents are given in Figure \ref{fig:fig6}. 
The exponents do not produce any significant differences, suggesting that the structural information that can be concluded are not particularly different in the different representations. This suggests that the structure is maintained in the rank-3 representation.

The only salient difference is that rank-4 allows specific affiliation pairs to be investigated albeit with issues in statistical significance if there are not enough data-points for the given number of affiliations. 
Rank-3 can easily identify the inter-affiliation links in a specific slice, and is thus well-suited to investigate inter-affiliation dynamics as a whole, whilst mitigating the issues of statistical significance. 

\begin{figure}
  \centering
  \includegraphics[width=0.50\textwidth]{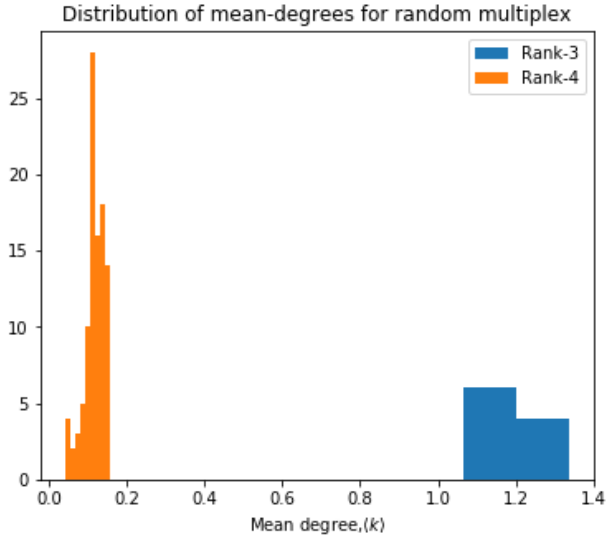}
  \caption{Histograms of the approximated exponents of the slice degree distributions for the rank-3 and rank-4 frameworks for randomly generated dataset. These only include trend lines for statistically significant regressions. The bins are selected based on the Freedman-Diaconis rule. The shapes and peak values do not differ significantly between the rank-3 and rank-4 frameworks.}
  \label{fig:figa}
\end{figure}

\textbf{For the Erdős–Rényi network}, the resulting degree distributions all form Poisson distribution in every slice. 
Due to the sparsity of the links, only the right-tail of the expected Poisson distribution appears (both tails appear at higher link probabilities e.g. $p=0.15$). 
The only differences between the two representations are the magnitudes of the counts and the value of mean-degree as shown in Figure \ref{fig:figa}. 
Given that all links and affiliations are randomly distributed, it is natural that the number of links is lower in the rank-4 representation. 
It is therefore expected that the $\langle k \rangle_{rank-4} \approx \frac{1}{M}\langle k \rangle_{rank-3}$. 

Ultimately, the overall information that can be concluded regarding structure remains the same between the rank-3 and rank-4 representations. 
It is only the magnitudes that change. 
The rank-4 can provide specificity, whilst the rank-3 framework has an advantage in statistical significance. 

\subsubsection{Node activity distribution}

\begin{figure}
  \centering
  \includegraphics[width=0.80\textwidth]{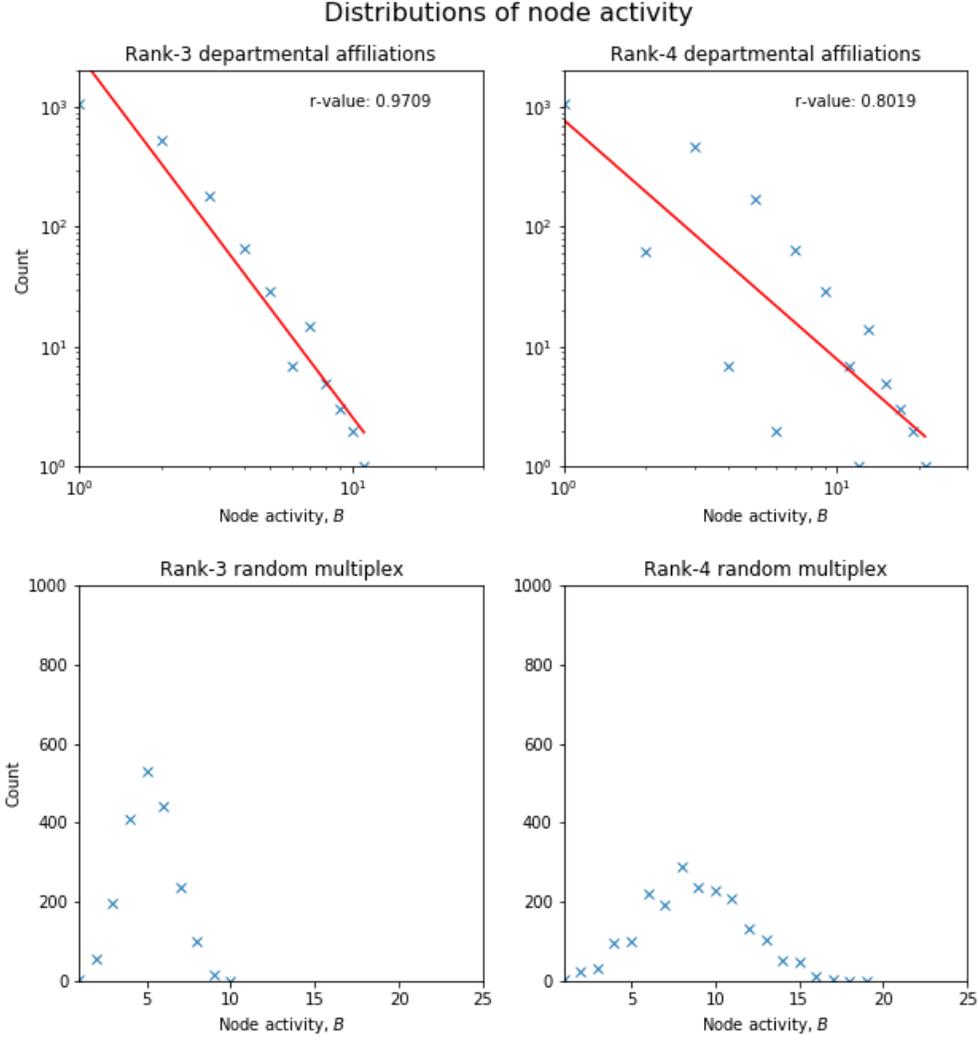}
  \caption{Node activity distributions for the rank-3 and rank-4 frameworks for the University of Bath co-authorship dataset. The rank-4 framework produces more noise in comparison to the rank-3 framework.}
  \label{fig:fig7}
\end{figure}

The node activity distribution provides information on how much presence a node has in multiple slices. 
The resulting distributions for both representations in the empirical and randomly generated datasets are shown in Figure \ref{fig:fig7}. 

\textbf{For the empirical dataset}, both representations are well approximated by a negative power-law relationship. 
The rank-3 representation has a correlation coefficient of 0.9709, whilst the rank-4 representation has a correlation coefficient of 0.8019. 
This can be partially explained by the fact that there are fewer slice in the rank-3 representation, and would thus naturally have less noise (i.e. has a higher Pearson correlation coefficient; see Figure \ref{fig:fig7}). 
No further information can be gained in the rank-4 abstraction over the rank-3 abstraction, meaning that the rank-3 representation is the most useful.

\textbf{For the Erdős–Rényi network}, node activity distribution for both representations are well approximated by a Poisson distribution. 
As with the empirical dataset, there is more noise in the rank-4 representation, whilst providing no further benefit.

\subsubsection{Slice-pair closeness distribution}

\begin{figure}
  \centering
  \includegraphics[width=0.80\textwidth]{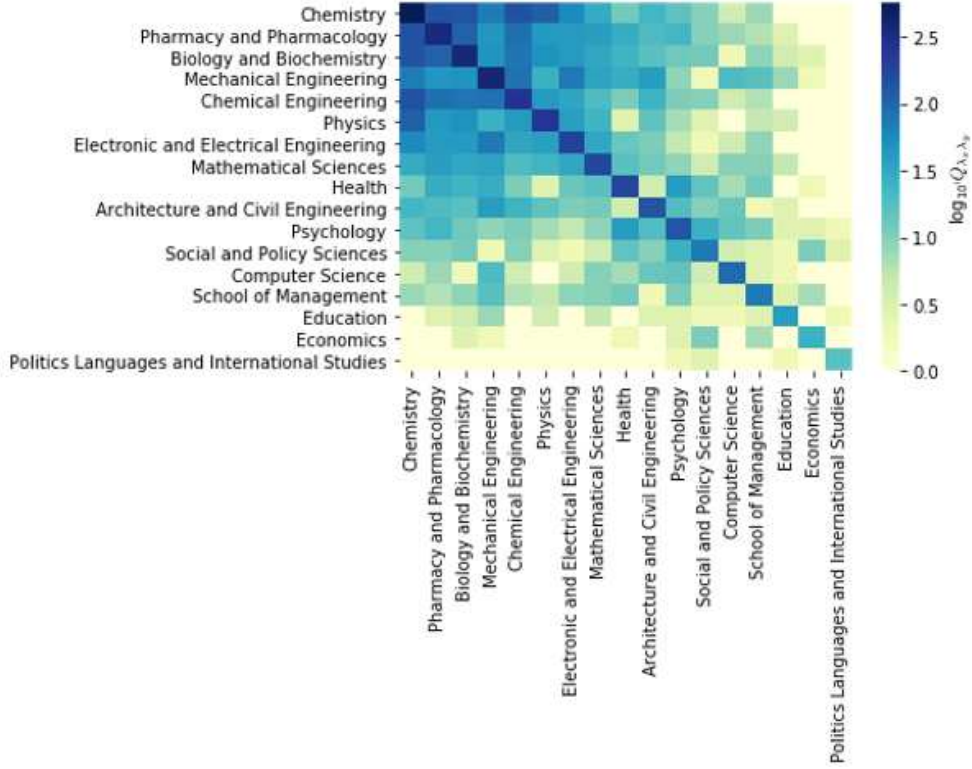}
  \caption{Heatmap of the slice-pair closeness for the rank-3 framework of the University of Bath co-authorship dataset. It exemplifies how similar affiliations are on a logarithmic scale.}
  \label{fig:fig_heatmap}
\end{figure}

The slice-pair closeness distribution provides a measure of how much similar acticity two slices have, and is exemplified in the heatmap of the empirical dataset shown in Figure \ref{fig:fig_heatmap}.  

\begin{figure}
  \centering
  \includegraphics[width=0.80\textwidth]{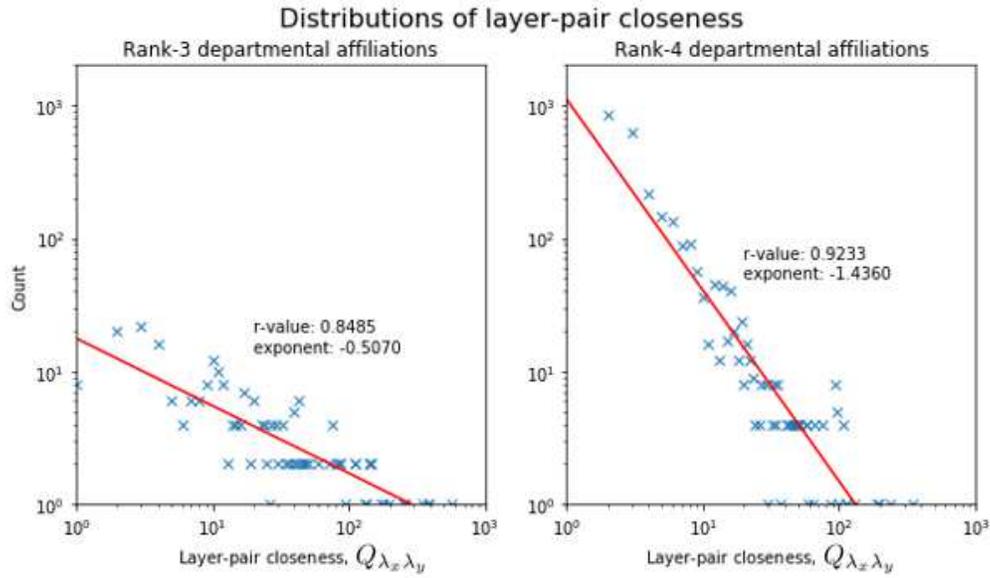}
  \caption{Distribution of the slice-pair closeness values for the rank-3 (left) and rank-4 (right) frameworks for the University of Bath co-authorship dataset.}
  \label{fig:fig8}
\end{figure}

\textbf{For the empirical dataset}, both representations provide a good fit with a negative exponent (discounting the values of $Q_{\lambda_{x}\lambda_{y}}$ that are zero) as shown in Figure \ref{fig:fig8}. 
This suggests that there are relatively few slices that are closely related, with the majority being poorly related. 
The biggest difference is in the magnitude of the exponent and the spread of distribution. 
The rank-3 is flatter, meaning that there are more slice-pairs with high closeness scores, and significantly fewer slice-pairs with low closeness scores. 
The conclusions that can be drawn from these distributions are similar, however as the rank-4 representation has a stronger regression, it has more confidence in the conclusions that can be drawn.

The slice-pair closeness centrality, $Q_{\lambda}$, follows this pattern. 
A negative exponent distribution is exhibited for the departmental affiliation data. 
However, the rank-3 can only create a distribution from $17$ data points, resulting in a distribution with points. 
Whilst this is well approximated by a power-law exponent, a second-order polynomial would perfectly fit the points as well. The rank-4 distribution creates a distribution from $17^{2}$ data points, and the resulting distribution well-approximated by a negative power-law exponent.

\begin{figure}
  \centering
  \includegraphics[width=0.80\textwidth]{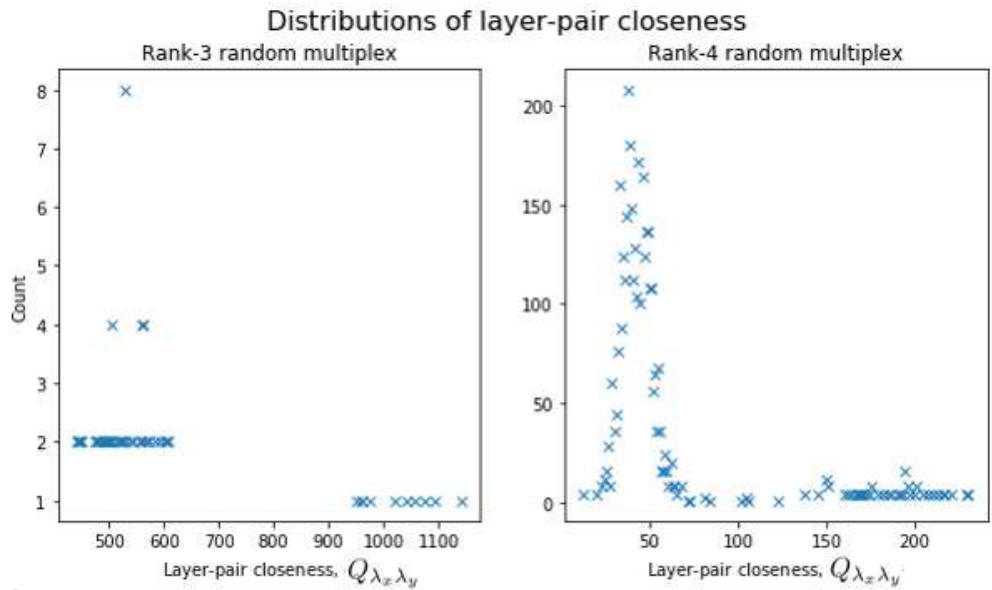}
  \caption{Distribution of the slice-pair closeness values for the rank-3 (left) and rank-4 (right) frameworks for the randomly generated dataset.}
  \label{fig:figb}
\end{figure}

\textbf{For the Erdős–Rényi network}, the slice-pair distributions exhibit a Poisson distribution for both the rank-3 and rank-4 frameworks as shown in Figure~\ref{fig:figb}. 
However, as with the departmental affiliation, the rank-3 distribution is noisier (R-value of $0.8485$ in comparison to $0.9233$ for the rank-4 representation). 
The noise makes it difficult to identify the Poisson distribution. 
The major difference here is the count magnitudes, with the highest count being only 8 for the rank-3 distribution. 

The slice-pair closeness centrality, $Q_{\lambda}$ exhibits similar issues as for the empirical dataset. 
Although even with the $10^{2}$ data points, the resulting Poisson distribution is not clearly visible. In experimentation with different parameters for the randomly generated network, the Poisson distribution emerges in much larger networks (e.g.\ with 20,000 nodes, $\sim$60,000 links, and 50 different affiliations). 

\begin{figure}
  \centering
  \includegraphics[width=0.80\textwidth]{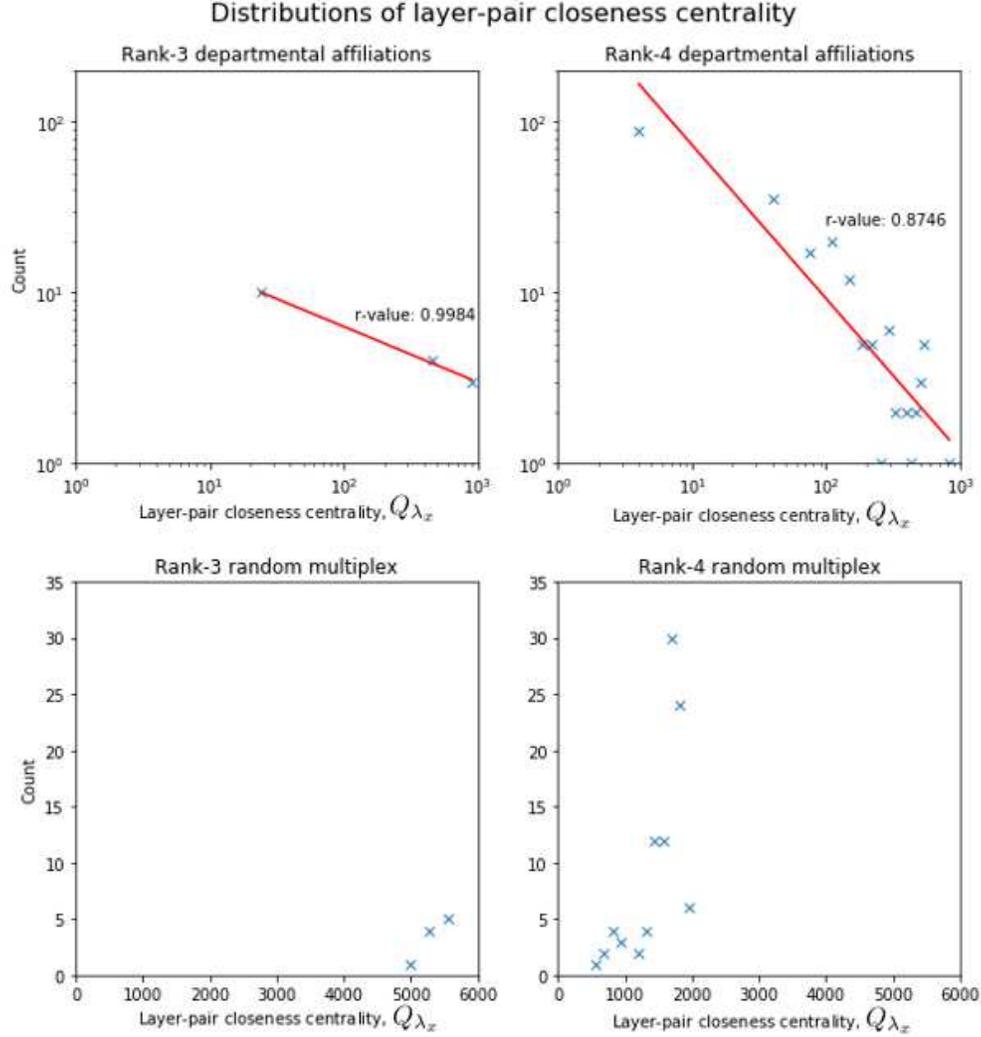}
  \caption{Distributions of the slice-pair closeness centrality.}
  \label{fig:fig9}
\end{figure}

The slice-pair closeness measure is an interesting case where a greater number of slices improves statistical significance due to the presence of more data points. This is because it is a slice-based measure.

\subsection{Discussion}

The results compare three metrics for the rank-3 and rank-4 representations across two datasets. 
This discussion considers the similarities, differences, benefits, and drawbacks of the rank-3 and rank-4 representations. 

The same type of relationships are exhibited across all metrics and datasets. 
For the node degree, node activity, and slice-pair closeness, the same type of relationship occurs for both the rank-4 and rank-3 representations. 
Some minor differences in mean-degree values can be seen in generated dataset. 
However, this can be almost entirely attributed to the fact that all nodes are present in every slice and both representations are showing the same information, dividing the mean-degree by a factor of $M$, which matches the findings shown in Figure 7. 
It is worth noting that the rank-3 representation has on average $M$ times more datapoints per slice, producing stronger evidence for a regression. 
The node activity produces very similar plots, showing a power-law regression in the empirical dataset and a Poisson distribution for the Erdős–Rényi network for both representations. 
The node activity is a node-based metric and will thus have the same number data points in the rank-3 and rank-4 representations. 
However, as a node can be active in more slices in the rank-4 representation, the results are more spread out and exhibit more noise. 
The slice closeness results show very similar distributions.
However, due to the small number of data points for the rank-3 representation (as there can only be $M$ data points), the distributions are either noisy or cannot establish trends.

Thus, several things have been shown. 
Firstly, using the metrics and datasets established, no significant differences can be found between the structures of the rank-4 and rank-3 representations. 
Thus, the rank-3 representation maintains the structure of the rank-4 representaion.
Secondly, the rank-3 representation regressions are on average $M$ times more densely populated with data points than its rank-4 counter part when the metric is node based. 
The opposite holds true for metrics that are slice-based. 

Finally, whilst this research has focused on the similarities of the rank-3 representation in comparison to the rank-4, it is important to highlight that the representations show different aggregations of the same system. If the focus on the research is on a specific interaction between two known affiliations, a rank-4 representation can more directly be used to investigate this. However, if the focus of the research is on intra- or inter-affiliation dynamics, the rank-3 representation is more appropriate. 

\section{Conclusion}

Single-affiliation systems are a sub-problem of affiliation systems that add constraints limiting the effectiveness of commonly used rank-4 multilayer network approaches. While they are able to capture the dynamics of the system, the resulting data structure is sparse, leading to challenges in generating statistically significant findings. This was observed in the analysis of a co-authorship dataset, where 72.3\% of slices could not produce statistically significant degree distributions (see Figure \ref{fig:fig5}).

This paper has overcome the limitations of rank-4 approaches by using a rank-3 representation for single-affiliation systems. The rank-3 is able to maintain all the information of a rank-4 representation for single-affiliation networks while increasing the statistical confidence of the findings. Ultimately, both frameworks are valid. However, the rank-3 representation provides significant benefits in allowing systems to be investigated with greater statistical confidence, whilst the rank-4 framework should be used when the specific interaction between two known affiliations needs to be investigated in detail.

\bibliographystyle{unsrt}  
\bibliography{references}

\end{document}